\documentclass[12pt]{article}

\usepackage{amsmath}
\usepackage{amssymb}
\usepackage{amsthm}

\usepackage{upref}

\newtheorem{lem}{Lemma}[section]
\newtheorem{thm}{Theorem}[section]
\newtheorem{cor}{Corollary}[section]

\newcommand\Cal[1]{{\cal #1}}

\newcommand\naturals{\mathbb N}
\newcommand\complex{\mathbb C}

\newcommand\Prob{{\rm Pr}}

\newcommand\w{{\rm wp}}
\newcommand\ex{{(q)}}
\newcommand\we{{\rm we}}
\newcommand\ap{{\rm ap}}
\newcommand\avex{{\rm ae}}
\newcommand\sym{{\rm sym}}
\newcommand\loss{{\rm loss}}
\newcommand\win{{\rm win}}

\def\d{{\delta}}
\def\D{{\Delta}}
\def\c{{\gamma}}

\def\e{{\varepsilon}}

\def\slug{\rule{0.1in}{0.1in}} 

\title{\bf Average case quantum lower bounds \\
for computing the boolean mean}
\author{A. Papageorgiou \\
Department of Computer Science \\
Columbia University \\
New York, NY 10027}

\date{June 2003}
\begin{document}

\setcounter{page}{1}
\maketitle

\begin{abstract} We study the average case approximation of the Boolean 
mean by quantum algorithms. We prove general query lower bounds for classes 
of probability measures on the set of inputs. 
We pay special attention to two probabilities, 
where we show specific query and error lower bounds and the 
algorithms that achieve them. 
We also study the worst expected
error and the average expected error of quantum algorithms and show 
the respective query lower bounds. Our results extend the optimality of 
the algorithm of Brassard et al. 

\end{abstract}

\vskip 2pc
\section{Introduction}
\vskip 1pc

Quantum computers can solve certain problems significantly faster than
classical computers.
One of these problems is the approximation of the mean of a Boolean function
or, equivalently, the approximation of the mean of $n$ Boolean variables.
Suppose that the input is presented as a black-box or an oracle,
which the algorithm queries \cite{beals}.
Classical algorithms require $\Theta( n(1-\e))$ evaluations
(or queries)
in the worst case, for error at most $\e$. Classical randomized algorithms
solve this problem faster by requiring $\Theta(\min \{ \e^{-2},n\})$
evaluations. Quantum algorithms solve the problem in the worst case
with high probability and are superior because they
require only $\Theta(\min \{ \e^{-1},n\})$ queries. 

More specifically, Brassard et al.\cite{brassard} exhibited
an algorithm achieving accuracy $\e$ with a number of 
queries proportional
to $\min \{ \e^{-1},n\}$.
This algorithm is based on 
Grover's quantum search algorithm; see \cite{nielsen} for
a description of Grover's algorithm and for details about quantum computing.
The lower bounds of Nayak and Wu \cite{nyakWu} 
establish the asymptotic optimality
of the algorithm of Brassard et al. in the worst case.

Instead of the worst case error, we can consider
the average error of quantum algorithms with respect to a probability
measure on the class of the inputs.
The average case is important for two reasons. The first one is that
it may reduce the query complexity. The second is that if we know
that the querey complexity is not reduced then the worst case results and
the optimality of known algorithms is extended.
It is also important to derive classes of measures 
for which similar complexity results hold.
In this paper we deal with these issues.

In particular, for the approximation of the mean
of $n$ Boolean variables with uniform distribution on the
set of inputs,
the average error of any quantum algorithm, with $T$ queries of order
$o(n)$,
is $\Omega( \min\{n^{-1/2}, T^{-1}\} )$. The query complexity
is zero as long as $\e$ is $\omega ( n^{-1/2})$
\footnote{$f(n)$ is $\omega(g(n))$ $\Leftrightarrow$ $g(n)$ is $o(f(n))$.}. 
When $\e=\Theta(n^{-1/2})$ the query complexity remains zero as long as
the asymptotic constant is large, but when this constant is small
the query complexity is $\Omega(n^{1/2})$.
The query complexity becomes
asymptotically equal to that of the worst case
when $\e$ is $o(n^{-1/2})$. 
On the other hand, if all possible values of the mean are uniformly distributed
then the average error of any algorithm is $\Omega( T^{-1})$.
In this case, the query complexity is asymptotically equal to that of the
worst case for all values of $\e$.

We generalize our results by showing conditions on classes of measures
under which
the query complexity is asymptotically equal to that of the worst 
case as long as $\e$ is appropriately small.
Our results extend the optimality of the quantum algorithm of 
Brassard et al. when high accuracy is important. 

Quantum algorithms are probabilistic in nature. 
For a given input, they can produce various outcomes, each with
a certain probability. Typically,
we want them to achieve a given accuracy with probability greater
than $\tfrac 12$ and, therefore, we study their probabilistic error.
On the other hand,
we can study the expected error of a quantum algorithm
by considering its average error with respect to all outcomes 
resulting from a given 
input. Therefore, for a class of inputs we study the worst expected error.
This is also an
intuitive error criterion and is similar to the way we measure
the error in Monte Carlo integration.
We show that any algorithm with worst expected error at most
$\e$ must make $\Omega( \e^{-1}, n)$ queries. Therefore, the algorithm of 
Brassard et al. with repetitions as described in \cite{hkw} 
is asymptotically optimal. 

We also show that the query lower bounds that hold for the average case
remain valid when we consider the average expected error of quantum algorithms.
In this case we consider a probability measure on the set of inputs, and
for each input we consider the expected error of the algorithm with
respect to all possible outcomes.

Finally, it is easy to see that an algorithm approximating the Boolean mean
can be used to approximately count the number of ones among
$n$ Boolean variables. Therefore, all our results directly extent to 
approximate counting
and we exhibit the corresponding query and error lower bounds.

\vskip 2pc
\section{Problem Definition}
\vskip 1pc

Let $B_n=\{ 0, 1\}^n$ denote all tuples of $n$ Boolean variables.
We assume that any $X=(x_1,\dots,x_n) \in B_n$ 
is given by an oracle or a black box, which
on input $i$ outputs $x_i$. Oracle access of this type is called a  
{\it query}.  
We want to compute the mean of $X$, i.e.,
\begin{equation*}
a_X=\frac {|X|}{n}, \quad {\rm with\ } |X|=\sum_{i=1}^n x_i.
\label{eq:meandef}
\end{equation*}

In this paper we consider the quantum query model of Beals 
et al.\cite{beals}, 
where the cost of an algorithm is the number of its queries.
A quantum algorithm applies a sequence of unitary 
transformations, which include queries, to an initial state, 
and at the end the final state is measured.
See 
\cite{beals,heinrich,nielsen}
for the details of the model of computation, which we summarize below 
to the extent necessary for this paper.

A quantum algorithm has the form
\begin{equation*}
U_TQ_XU_{T-1}Q_X\cdots U_1Q_XU_0 |\psi_0\rangle=: |\psi\rangle,
\end{equation*}
where $U_0,\dots, U_T$
are unitary transformations that do not depend on the input
$X$, 
the operator $Q_X$ is also a unitary transformation and corresponds to a query
to the oracle, the integer 
$T$ is the number of times $Q_X$ is applied, that is 
the number of queries, 
$|\psi_0\rangle$ is the initial state on which the sequence of 
transformations is applied, and $|\psi \rangle$ is the final state
of the algorithm which is measured.
The states $|\psi_0\rangle$ and $|\psi\rangle$ are unit vactors of
$\Cal H_m=\underbrace{\complex^2\otimes \cdots\otimes \complex^2}_m$,
for some appropriately chosen $m\in\naturals$.
The measurement produces 
one of $M$ outcomes. Outcome $j\in\{0,\dots,M-1\}$ occurs with
probability $p_X(j)$, which depends on $j$ and the input $X$.

In principle, quantum algorithms may have many measurements
applied between sequences
of unitary transformations of the form above. However, any
algorithm with many measurements and a total of $T$ queries can be simulated by
an algorithm with only one measurement that has $2T$ queries
\cite{heinrich}. Hence, without loss of generality we consider 
the cost of algorithms with a single measurement.

Given an outcome $j$, 
we approximate $a_X$ by a number $\hat a_X(j)$. 
Note that $\hat a_X(j)$ depends on the input $X$ and the outcome of 
the measurement.
Given a probability $p> \frac 12$, the error of a quantum algorithm 
with $T$ queries on input $X$ is defined by
\begin{equation*}
e(X,T,p)=\inf \left\{ \c :\quad \sum_{j:|a_X-\hat a_X(j)|\le \c}
p_X(j)\geq p\right\}.
\label{eq:localperr}
\end{equation*}

The {\it worst probabilistic} error of a quantum algorithm with $T$ queries 
in the class $B_n$ is defined by
\begin{equation*}
e^{\w}(B_n,T,p)=\max_{X\in B_n}e(X,T,p).
\label{eq:wperr}
\end{equation*}
As we have mentioned,
Brassard et al.\cite{brassard} show a quantum
summation algorithm (QS) for computing the Boolean mean
and study its properties using this error criterion. 
The query lower bound $\Omega(\min(\e^{-1}, n))$ of
Nayak and Wu \cite{nyakWu} also holds in the worst case.

In this paper we consider the {\it average probabilistic} error of a quantum
algorithm in the class $B_n$ which we define by
\begin{equation*}
e^{\ap}(B_n,T,p)=\sum_{X\in B_n}e(X,T,p) \mu(X),
\label{eq:aperr}
\end{equation*}
where $\mu$ is a probability measure on the set of inputs $B_n$.

In the next section we exhibit conditions for classes of probability measures 
and prove the corresponding query lower bounds. We will pay special attention
to the following two measures
\begin{eqnarray*}
\mu_1(X)&=& 2^{-n}, \quad \forall\;  X\in B_n \\
\mu_2(X)&=&  \frac 1{(n+1){n \choose k}}, \quad {\rm for\ } X\in B_n 
{\rm\ with\ } |X|=k.
\end{eqnarray*}
The first measure corresponds to
the case where all inputs are equally likely, while the second 
measure corresponds to the case where all possible values of the
mean are equally likely. 

We now define the {\it worst expected} error of a quantum algorithm with
$T$ queries in the class $B_n$ as
\begin{eqnarray*}
e^{\ex}(X,T) &=& \left\{ \sum_{j=0}^{M-1} 
|a_X-\hat a_X(j)|^q p_X(j) \right\}^{1/q} \\
e^{\we}(q,B_n,T)&=&
\max_{X\in B_n} e^{\ex}(X,T), \quad {\rm\ with\ }  1\le q<\infty,
\end{eqnarray*}
where the summation is over all possible outcomes.
Note that in this case
we consider all the outcomes and not just outcomes that occur
with probability $p> \tfrac 12$.

We also consider the average with respect to the inputs of the expected
error, with respect to the outcomes, of a quantum algorithm with $T$ queries
in the class $B_n$. We call this the {\it average expected} error and we 
define it by
\begin{equation*}
e^{\avex}(q, B_n, T)= \sum_{X\in B_n} e^{\ex}(X,T)\mu(X), \quad {\rm\ with\ }
1\leq q < \infty,
\end{equation*}
where $\mu$ is a probability measure on the set of inputs $B_n$.

Finally,
Nayak and Wu \cite{nyakWu} show query lower bounds for a number of different 
problems. 
One of them is the computation of a
{\it $\D$-approximate count}, i.e., a number $\hat t_X$ such that
$|t_x -\hat t_X|<\D$, where $t_X=|X|=n\cdot a_X$ for $X\in B_n$.
This problem is directly related to the approximation of the Boolean mean.
We study it on the average by appropriately defining the error of a
quantum algorithm. In particular, we set
\begin{eqnarray*}
e_1(X,T,p) &=& n \cdot e(X,T,p) \\
e^{\ap}_1(B_n,T,p) &=& n \cdot e^{\ap}(B_n, T, p) \\
e^{\we}_1(q,B_n, T) &=& n\cdot e^{\we}(q,B_n,T), \quad {\rm\ with\ }
1\le q< \infty, \\
e^{\avex}_1(q,B_n,T) &=& n \cdot e^{\avex}(q, B_n, T),\quad {\rm\ with\ }
1\le q< \infty.
\end{eqnarray*}
Our results concerning the mean directly extend to approximate counting 
by setting $\Delta = n\e$.

\vskip 2pc
\section{Average probabilistic error}
\vskip 1pc

Kwas and Wo\'zniakowski \cite{kwas} show that the QS algorithm has zero worst 
probabilistic error when
the number of its queries is greater than $\tfrac 32 \pi n $,
for $p\le \tfrac{8}{\pi^2}$. Trivially, QS
also has zero average probabilistic error in that case.
Therefore, we study the error of quantum algorithms when the
number of their queries is of order $o(n)$.

It is convenient to deal with arbitrary measures first and then use
the results in the study of $\mu_1$ and $\mu_2$.
So we begin by defining classes of probability measures
and deriving the  corresponding query lower bounds. 
Roughly speaking, all the measures $\mu$ in a 
class satisfy the same lower bound for $\mu(X)$ as long $|X|$ belongs to a
certain subset $I$ of $\{ 0,\dots, n\}$. This lower bound depends on $|X|$,
on $I$ and, particularly, its cardinality $|I|$. 
For example, for the set
$I=\{ \tfrac n2 - \sqrt{n}, \dots,  \tfrac n2 + \sqrt{n}\}$, which has
cardinality
$|I|= 2\sqrt{n} + 1$, we can use  $c [\sqrt{n} 
{n\choose |X|}]^{-1}$, (where $c>0$ is a constant and $|X|\in I$,) 
as the lower bound defining the class 
of measures. Observe, that due to Lemma 6.1 in Appendix $\mu_1$
asymptotically satisfies this lower bound on $I$.
Similarly, for the set  
$I=\{ \tfrac n4,\dots, \tfrac {3n}4 \}$, which has cardinality 
$|I|=\tfrac n2 +1$,
we can use $c[ n {n \choose |X|}]^{-1}$ (where $c>0$, 
is a constant and $|X|\in I$) as the lower bound
in the definition of the class of measures. Note that $\mu_2$ asymptotically
satisfies this lower bound on $I$.
Clearly these two choices of $I$ distinguish two classes
of probability measures. The cardinality of the set $I$ as a function
of $n$ is important in our analysis.
We will assume that $|I|\to \infty$ as $n\to \infty$ as in the previous two 
examples. We also consider the number of queries $T$ and the desired accuracy
$\e$ as functions of $n$ and carry out an asymptotic analysis. In this 
and in the following sections the implied asymptotic constants are absolute
constants.

\begin{thm} 
Consider the approximation of the Boolean mean.
Let $I\subseteq \{0,\dots,n\}$ be a set of consecutive indices, such that
its cardinality $|I|$ is $\omega(1)$ as a function of $n$. Assume that
$k(n-k)$ is $\Theta(n^2)$ for every $k\in I$. Let $\mu$ be a
probability measure on $B_n$ such that 
$$\mu(X)=\Omega(|I|^{-1})\frac 1{{n\choose |X|}}, \quad
{\rm for\ every\ }  X\in B_n {\rm \ with\ }|X|\in I.$$
Then for any $\e>0$ of order $o(|I|n^{-1})$, the condition 
$e^{\ap}(B_n,T,p)\le\e$
implies that $T$ must be $\Omega(\min(\e^{-1}, n))$.
\end{thm}

\noindent {\bf Proof:} 
We will prove the Theorem for $\e\ge 1/n$. The case $\e <1/n$ will
then follow immediately. 

Consider a quantum algorithm with $T$ queries that has error
$e^{\ap}(B_n,T,P)\le\e$.
Using the lower bound on $\mu(X)$ in the assumption of this theorem, we have
$$
\e\geq \sum_{X\in B_n}e(X,T,p)\mu(X)\geq \frac c{|I|}
\sum_{k\in I}\frac{1}{{n \choose k}}\sum_{X:|X|=k}e(X,T,p),
$$
where $c>0$ is a constant.

We multiply both sides of the inequality by $n$ and define
$\Delta = c^{-1}n\e$ and $\Delta(k)=n \sum_{|X|=k}e(X,T,p)/{n\choose k}$ 
and use the Markov inequality to obtain
\begin{eqnarray*}
\Delta &\ge& \frac 1{|I|} \sum_{k\in I} \Delta(k) \\
&=& \frac 1{|I|} \left\{ \sum_{k:\;\Delta(k)< 2\Delta } \Delta(k)
+ \sum_{k:\;\Delta(k)\ge 2\Delta } \Delta(k) \right\} \\
&\ge & \frac 1{|I|} 2\Delta n_+,
\end{eqnarray*}
where $n_+$ is the number of indices for which $\Delta(k)\ge 2\Delta$.
Clearly $n_+\le \tfrac 12 |I|$ and, therefore, 
$n_-:=|I| -n_+ \ge\tfrac 12 |I|$.
Thus for at least half of the indices in $I$ we have
$\Delta(k) < 2\Delta$. We define $J$ to be the set of all these indices. 
Note that $\Delta$ is $o(|I|)$ because $\e$ is $o(|I|n^{-1})$.
Without loss of generality we assume that
$\Delta$ is an integer, since otherwise we can replace it by its ceiling, which
does not change its order of magnitude.

Now consider $k\in J$ so that $\Delta(k)< 2\Delta$ and let
$\d(X,k)=n\, e(X,T,p)$ for $|X|=k$. Then for $m\ge 2$, 
which we will further specify later,
we have
\begin{eqnarray*}
2\Delta &>& \frac 1{{n \choose k}} \sum_{|X|=k} \d(X,k) \\
&=& \frac 1{{n \choose k}}
\left\{ \sum_{|X|=k:\; \d(X,k)< 2m\Delta}  \d(X,k) +
\sum_{|X|=k:\; \d(X,k)\ge  2m\Delta}  \d(X,k) \right\} \\
&\ge&  \frac 1{{n \choose k}} 2m\Delta \tilde n_+,
\end{eqnarray*}
where $\tilde n_+$ is the number of strings $X$ with $|X|=k$, for which
$\d(X,k)\ge 2m\Delta$. Clearly $\tilde n_+ \le {n\choose k}m^{-1}$ and,
therefore, $\tilde n_- :={n\choose k}- \tilde n_+ 
\ge (1-m^{-1}){n\choose k}$.
Thus for at least $\tilde n_-$ many strings $X$ we have 
$\d(X,T,p):=\d(X,k) < 2m\Delta$.

Since $|I|$ is $\omega(1)$ we claim that for sufficiently large $n$
there exist $k_1, k_2\in J$ 
that are at least $4m\Delta$ 
apart whose distance does not exceed $O(\Delta)$, i.e.,
$|k_1-k_2|\ge 4m\Delta$ and $|k_1-k_2|$ is $O(\Delta)$. 
Indeed, if we assume otherwise, we have
that $|k_1-k_2|< 4m\Delta$ or $|k_1-k_2|$ is $\omega(\Delta)$.
Consider the indices in $I$ in ascending order. Let $i_1\in I$ be the first
index that belongs to the set $J$ (recall that at least half of the indices
in $I$ belong to $J$). 
Based on our assumption, 
the next index (greater than $i_1$) that belongs to $J$
is either at a distance less than $4m\D$ away from $i_1$ 
or at a distance $\omega(\D)$
away from $i_1$. We group together all the indices that are at a distance
less than $4m\D$ away from $i_1$. Clearly there are no more 
than $4m\D$ indices in the group.
Now we consider the first index $i_2\in I$ that belongs to $J$ and is 
at a distance $\omega(\D)$ away from $i_1$. We repeat the same procedure 
using $i_2$ in the place of $i_1$, and we form a second group of indices that 
belong to $J$ and are at a distance less than $4m\D$ away from $i_2$.
As we iterate this procedure we form groups of indices
that belong to $J$, where each group is at a distance $\omega(\D)$
away from the group before it. 
We stop when
we exhaust the indices in $J$. It is clear that between every two 
groups we have
$\omega(\D)$ elements of $I$ that do not belong to $J$. It is also clear
that we have 
to repeat the above procedure at least $|J|/(4m\D)\ge |I|/(8m\D)$ times
in order to exhaust the indices in $J$. Considering the indices between the 
consecutive groups that do not belong to $J$ we conclude that the 
cardinality of $I$ must be at least $\omega(\D) [|I|/(8m\D)]=
\omega(|I|)$, which
is a contradiction.

Now we use the algorithm that approximates the mean to derive another
algorithm that approximates the partial Boolean function
\begin{equation*}
f_{k_1,k_2}(X)=\left\{ 
\begin{array}{lcl}
1 &\quad &{\rm if\ } |X|=k_1 \\
0 &\quad &{\rm if\ } |X|=k_2,
\end{array}
\right.
\end{equation*}
where, without loss of generality, we can assume that $k_1 > k_2$.

The description of the new algorithm $A$ is as follows: 
On input $X$, where $|X|=k_1$ or $k_2$, we run
the algorithm that approximates the mean and if the value of the result
$\hat a_X(j)$ satisfies $|k_1-n\hat a_X(j)|<2m\Delta$ then the new algorithm
outputs $1$. It outputs $0$ otherwise.

Let's look at the success probability $\Prob\{ A(X)=f_{k_1,k_2}(X)\}$
of the new algorithm 
for the different inputs for which $f_{k_1,k_2}$ is defined. 
If $|X|=k_1$ we have
\begin{eqnarray*}
\sum_{|X|=k_1} \Prob\{ A(X)=f_{k_1,k_2}(X)\}
&=& \sum_{|X|=k_1} \Prob\{ A(X)=1 \} \\
&\ge&  \sum_{|X|=k_1,\; \d(X,T,p) < 2m\Delta} \Prob\{ A(X)=1 \} \\
&=& \sum_{|X|=k_1,\; \d(X,T,p)<2m\Delta} \Prob\{ |k_1-n\hat a_X(j)|<2m\Delta \}\\
&\ge& (1-m^{-1}) {n \choose k_1} p,
\end{eqnarray*}
because $\d(X,T,p)<2m\Delta$ is equivalent to $e(X,T,p)< 2m\Delta n^{-1}$, 
which implies that $|k_1/n -\hat a_X(j)|< 2m\Delta n^{-1}$ holds with  
with probability
at least $p$. The fact that $\tilde n_-\ge (1-m^{-1}) 
{n \choose k_1}$ yields the final inequality.
Therefore, the probability that algorithm $A$ fails on any input $X$
for which $|X|=k_1$ satisfies
\begin{equation}
\sum_{|X|=k_1}\Prob \{ A(X)\ne f_{k_1,k_2}(X) \} = 
\sum_{|X|=k_1}\Prob \{ A(X)=0\} \le {n\choose k_1} \left(1-p+\frac pm\right).
\label{eq:probloss1}
\end{equation}
Let $c_2=1-p+p/m$. 
We choose $m$ in a way that
that $c_2< \tfrac 12$.

From \cite{beals} we know that the acceptance probability $q(X)=
\Prob\{ A(X)=1\}$ of a quantum algorithm $A$ is a real multilinear 
polynomial of 
degree at most $2T$, where $T$ is the number of its queries. 
Recall that 
the symmetrization of $q$ is the polynomial
\begin{equation}
q^{\sym}(X)=\frac {\sum_{\pi} q(x_{\pi(1)},\dots,x_{\pi(n)} )}{n!}, \quad
X=(x_1,\dots,x_n)\in B_n,
\label{eq:qsym}
\end{equation}
where the sum is over all permutations of the integers $1,\dots, n$.
Minsky and Papert \cite{minsky} show that there is a representation of
$q^{\sym}$ as a univariate polynomial in $|X|$ of degree at most that
of $q^{\sym}$.
For simplicity, with a slight abuse of notation we denote this univariate 
polynomial using the same symbol, i.e., $q^{\sym}(|X|)$.

In particular, for $|X|=k_1$ we have
$q(X)=1-\Prob\{ A(X)=0\}$, which implies $\Prob\{ A(X)=0\} = 1-q(X)=
f_{k_1,k_2}(X)-q(X)$. Thus
\begin{eqnarray*}
\sum_{|X|=k_1}\Prob\{ A(X)=0\}&=& {n\choose k_1}-\sum_{|X|=k_1}q(X) \\
&=& {n\choose k_1} - {n\choose k_1} q^{\sym}(|Y|) \\
&=& {n\choose k_1} \left[ f_{k_1,k_2}(Y) - q^{\sym}(|Y|) \right],
\quad\forall\; |Y|=k_1,\; Y\in B_n.
\end{eqnarray*}
The second equality holds because
when $X$ has $k_1$ ones in particular locations and $n-k_1$ zeros in the 
remaining locations then the $k_1!(n-k_1)!$ permutations of $X$ (when only the 
$k_1$ ones or the $n-k_1$ zeros are permuted) yield tuples
that are identical to $X$. 
Therefore, every term in $\sum_{|X|=k_1} q(X)$ appears $k_1!(n-k_1)!$ times
in the $\sum_{\pi} q(x_{\pi(1)},\dots,x_{\pi(n)} )$ of all permutations.
Thus considering (\ref{eq:qsym}) we have
$\sum_{\pi}q(x_{\pi(1)},\dots,x_{\pi(n)}) = k_1!(n-k_1)!
\sum_{|X|=k_1}q(X)$. 

Using (\ref{eq:probloss1}) and the last equality concerning the probability of 
failure of $A$ we obtain
\begin{equation}
c_2 \ge f_{k_1,k_2}(X)-q^{\sym}(|X|), \quad\forall\; |X|=k_1.
\label{eq:psymest1}
\end{equation}

We work similarly when $|X|=k_2$. We have
\begin{eqnarray*}
\sum_{|X|=k_2} \Prob\{ A(X)=f_{k_1,k_2}(X)\}
&=& \sum_{|X|=k_2} \Prob\{ A(X)=0 \} \\
&\ge &  \sum_{|X|=k_2,\; \d(X,T,p) < 2m\Delta} \Prob\{ A(X)=0 \} \\
&=& \sum_{|X|=k_2,\; \d(X,T,p)< 2m\Delta} 
\Prob\{ |k_1-n\hat a_X(j)| \ge 2m\Delta \} \\
&\ge& \sum_{|X|=k_2,\; \d(X,T,p)< 2m\Delta} 
\Prob\{ |k_2-n\hat a_X(j)| < 2m\Delta \} \\
&\ge& \left( 1-\frac 1m \right) {n \choose k_2} p,
\end{eqnarray*}
because $\d(X,T,p)<2m\Delta$ is equivalent to $e(X,T,p)< 2m\Delta n^{-1}$,
which implies that $|k_2/n -\hat a_X(j)|< 2m\Delta n^{-1}$ holds 
with probability
at least $p$. The fact that $\tilde n_- \ge (1-m^{-1}) 
{n \choose k_2}$ yields the final inequality.
Therefore, the probability that algorithm $A$ fails on any input $X$
for which $|X|=k_2$ satisfies
\begin{equation}
\sum_{|X|=k_2}\Prob \{A(X) \ne f_{k_1,k_2}(X) \} =
\sum_{|X|=k_2}\Prob \{ A(X)=1\} \le {n\choose k_2} \left( 1-p+\frac pm\right).
\label{eq:probloss2}
\end{equation}
In terms of $q(X)$ and its 
symmetrization the last inequality becomes
\begin{equation*}
 {n\choose k_2} c_2 \ge \sum_{|X|=k_2}q(X)={n\choose k_2}q^{\sym}(|Y|),
\quad\forall\; |Y|=k_2,\; Y\in B_n,
\end{equation*}
where the inequality is obtained from (\ref{eq:probloss2}) with $c_2=1-p+p/m$,
and the equality holds for the same reasons 
as those concerning the permutations of only ones or zeros in $X$
which we explained before.
This implies
\begin{equation}
c_2 \ge q^{\sym}(|Y|)=|f_{k_1,k_2}(Y)-q^{\sym}(|Y|)|, \quad\forall\; |Y|=k_2.
\label{eq:psymest2}
\end{equation}

We combine (\ref{eq:psymest1}) and (\ref{eq:psymest2}) to obtain 
$$|f_{k_1,k_2}(X)-q^{\sym}(|X|)| \le c_2< \tfrac 12,$$
for all the $X$ for which this partial Boolean function is defined. 
Recall that symmetrization does not increase the degree of a polynomial,
which implies that $2T$ is greater than or equal to the degree of $q^{\sym}$.
Using
the results of Nayak and Wu \cite{nyakWu} concerning lower bounds
for the degree of polynomials approximating the partial Boolean
function $f_{k_1,k_2}$,  and our assumption that
$k(n-k)=\Theta(n^2)$, for all $k\in I$, we obtain that the
degree of $q^{\sym}$ is
$$\Omega\left( \sqrt{\frac n{|k_1-k_2|}} + 
\frac {\sqrt{\kappa(n-\kappa)}}{|k_1-k_2|}\right),$$
where $\kappa \in\{k_1,k_2 \}$ which maximizes $|\tfrac n2 -\kappa|$.
Therefore, the number of queries of the original algorithm is
$\Omega(n\Delta^{-1})$, which, in turn, is $\Omega(\e^{-1})$. \hfill\slug 

\vskip 1pc

Theorem 3.1 extends the optimality properties of QS to the average 
probabilistic case when high accuracy is important. 
It shows that QS is asymptotically optimal in computing the Boolean mean
as long as $\mu$ satisfies certain properties. The range of possible values
of $\e$ has to be appropriately small 
and this depends on the class of measures
through the cardinality of the set $I$.
The larger this set is the larger the range of $\e$
for which Theorem 3.1 holds and QS is asymptotically optimal. 
On the other hand, as we are about to see, 
when there is demand for relatively low accuracy there
can be other algorithms faster than QS.

Let us now consider $\mu_1$ where
all elements $X\in B_n$ are equally likely
having probability $2^{-n}$. Kwas and Wo\'zniakowski \cite{kwas} show
that, with probability $p=1$,
the algorithm that outputs $\tfrac 12$ on any input without any
queries at all, i.e., $T=0$, has error
\begin{equation}
e^{\ap}(B_n,0,1)= (2\pi n)^{-1/2}(1+o(1)).
\label{eq:constalgerr}
\end{equation}
However, reducing the error further requires $\Omega(n^{-1/2})$ queries, as
we see below.

\begin{lem}
Consider the measure $\mu_1$. There exists a constant $c>0$ such that
the condition $e^{\ap}(B_n,T,p)\le c n^{-1/2}$, $p> \tfrac 12$, 
implies that $T$ is $\Omega(n^{1/2})$.
\end{lem}

\noindent {\bf Proof:} 
The proof is very similar to that of Theorem 3.1.
We point out the differences
and we refer to the proof of Theorem 3.1 for the identical parts.

Recall that in the proof of Theorem 3.1 
equation (\ref{eq:probloss1}) lead us to select $m\ge 2$ such that 
$1-p+p/m< \tfrac 12$.
Consider any such $m$ here.

We set $c=e^{-6(m+1)^2-2}(2\pi)^{-1/2}$. We consider the sets
$I_1=\{ \tfrac n2 +m\sqrt{n},\dots, \tfrac n2 +(m+1)\sqrt{n}\}$ and 
$I_2=\{ \tfrac n2 -(m+1)\sqrt{n},\dots, \tfrac n2 -m\sqrt{n}\}$. 
Note that for the indices $k\in I_1\cup I_2$ we have $k(n-k)=\Theta(n^2)$.

Assume that
$$
cn^{-1/2}\ge \sum_{X\in B_n} e(X,T,p)\mu_1(X)\ge 
2^{-n}\sum_{|X|\in I_j} e(X,T,p), \quad j=1,2,$$
since $\mu_1(X)=2^{-n}$, for every $X\in B_n$.
From Lemma 6.1, in the Appendix, we have that 
${n \choose {|X|}}2^{-n}> c n^{-1/2}$,
when $|X|\in I_1\cup I_2$, $X\in B_n$. 
We multiply by $n$ both sides of the inequality above, and define
$\Delta = c n^{1/2}$ and $\Delta(k)=n\sum_{|X|=k} e(X,T,p)/
{n\choose k}$ to obtain
$$\Delta > n^{-1/2}\sum_{k\in I_j}\Delta(k), \quad j=1,2.$$
Thus, there exist $k_j\in I_j$, such that
$\Delta(k_j)<\Delta$, $j=1,2$.
Let $\d(X,k_j)=n\, e(X,T,p)$, $|X|=k_j$, $j=1,2$. Then we have
\begin{eqnarray*}
\Delta &>& \frac 1{{n\choose k_j}} \sum_{|X|=k_j} \d(X,k_j) \\
&=& \frac 1{{n\choose k_j}} \left\{ 
\sum_{|X|=k_j:\; \d(X,k_j)<m\Delta} \d(X,k_j) +
\sum_{|X|=k_j:\; \d(X,k_j)\ge m\Delta} \d(X,k_j) \right\} \\
&\ge & \frac 1{{n\choose k_j}} m\Delta \tilde n_{j,+}, \quad j=1,2,
\end{eqnarray*}
where $\tilde n_{j,+}$ is the number of strings $X$ with $|X|=k_j$, for which 
$\d(X,k_j)\ge m\Delta$. Just like in the proof of Theorem 3.1 we
conclude that the number of strings $X$ for which $\d(X,k_j)<m \Delta$ 
satisfies
$\tilde n_{j,-}\ge (1-m^{-1}){n\choose k_j}$, $j=1,2$.

Now we use the algorithm that approximates the mean to derive another 
algorithm that approximates the partial Boolean function
\begin{equation*}
f_{k_1,k_2}(X)=\left\{ 
\begin{array}{lcl}
1 &\quad &{\rm if\ } |X|=k_1 \\
0 &\quad &{\rm if\ } |X|=k_2.
\end{array}
\right.
\end{equation*}

From this point on the proof is identical to the proof of Theorem 3.1.
and we omit the details. The conclusion is that the algorithm that
approximates $f_{k_1,k_2}$ and, therefore, the original algorithm
must make $\Omega(n/\Delta)$ or, equivalently,
$\Omega(n^{1/2})$ queries. \hfill \slug

\vskip 1pc

Thus we need to study the error
of the algorithm when the number of queries is $\Omega(n^{1/2})$.
The following theorem deals with this case and also summarizes our results 
with respect to $\mu_1$.

\begin{thm} 
Consider that approximation of the Boolean mean
and the average probabilistic error of a quantum 
algorithm with respect to $\mu_1$. The following two statements hold.
\begin{enumerate}
\item Let $T$ be $o(n)$. Then the error of any quantum algorithm 
with $T$ queries satisfies
$$e^{\ap}(B_n,T,p) = \Omega\left( \min\{ n^{-1/2}, T^{-1}\}\right).$$
\item Let $\e>0$ be $o(n^{-1/2})$. Then
the number of queries $T(\e)$ for error at most $\e$
satisfies
$$T(\e)=\Omega\left(\min\{\e^{-1}, n\}\right).$$
\end{enumerate}
\end{thm}

\noindent {\bf Proof:}  
The second statement directly follows from Theorem 3.1. Indeed, in
the proof of Lemma 3.1 we saw a lower bound for $\mu_1(X)$
when
$|X|$ belongs to sets of $n^{1/2}$ many indices close to $n/2$
(sets like $I_1$ and $I_2$). Thus the conditions of Theorem 3.1
hold for $\mu_1$ and the query lower bound is immediate.

Now we prove the first statement. From Lemma 3.1 we know that 
error less than $cn^{-1/2}$ requires 
$\Omega(n^{1/2})$ queries.
Hence, when the number of queries is 
$o(n^{1/2})$ then the error is bounded
from below by a quantity proportional to $n^{-1/2}$.

Let us now consider $T$ to be $\Omega(n^{1/2})$. Then $T^{-1}$ is
$O(n^{-1/2})$ and $\min\{ n^{-1/2}, T^{-1}\} = \Theta(T^{-1})$.
We prove the first statement by contradiction.  
Assume that $n$ is sufficiently large.
Suppose that the error lower bound is not 
$\Omega\left( \min\{ n^{-1/2}, T^{-1}\}\right)$ 
but that  $e^{\ap}(B_n,T,p)\le (Tg(T))^{-1}$, where $g$ is a
function such that $g(T)=\omega(1)$.
Set $\e=(Tg(T))^{-1}$ and observe
that $\e$ is $o(n^{-1/2})$. Then use the second statement of this theorem 
conclude that $T$ must be  
$\Omega(Tg(T))$, which is
a contradiction. \hfill\slug

\vskip 1pc

Kwas and Wo\'zniakowski \cite{kwas} show that for $\mu_1$,
the average probabilistic error of QS is $O(\min\{ n^{-1/2}, T^{-1} \})$
when the number of its queries is divisible by four.
Using Theorem 3.2 we conclude: 
\begin{itemize}
\item QS is an asymptotically optimal error algorithm.
\item QS makes an asymptotically optimal number of queries for accuracy $\e$,
when $\e$ is $\omega(n^{-1/2})$.
\item QS requires at least four queries
for error $O(n^{-1/2})$ when $\e$ is $\omega(n^{-1/2})$, 
while the optimal number of queries is zero, and is achieved 
by a constant algorithm.
\end{itemize}

We now consider $\mu_2$ which corresponds to the case that all values
of the mean are equally likely. As we shall see,
computing the mean in the average probabilistic case with $\mu_2$ is
just as hard as computing the mean in the worst probabilistic case.

\begin{thm}
Consider the approximation of the Boolean mean
and the average probabilistic error of a quantum 
algorithm with respect to $\mu_2$. The following two statements hold.
\begin{enumerate}
\item Let $T$ be $o(n)$. Then the error of any quantum algorithm 
with $T$ queries satisfies
$$e^{\ap}(B_n,T,p) = \Omega\left( T^{-1}\right).$$
\item Let $\e>0$ be $o(1)$.
Then the number of queries $T(\e)$ for error at most $\e$
satisfies
$$T(\e)=\Omega\left(\min\{\e^{-1}, n\}\right).$$
\end{enumerate}
\end{thm}

\noindent {\bf Proof:}
Trivially $\mu_2$ satisfies the conditions of Theorem 3.1 
for a set $I$ of $\Theta(n)$ many consecutive indices, e.g.,
  $I= \{ \tfrac n4,\dots, \tfrac 34 n \}$.
Therefore, the second statement is immediate.

We show the first statement by contradiction.
If $T$ is $O(1)$ then the error is bounded from below by a constant.
Indeed, if we assume
that $e^{\ap}(B_n,T,p)\le 1/g(n)$ for some function $g$ satisfying
$g(n)=\omega (1)$, then Theorem 3.1 yields that $T=\Omega(g(n))$, which is
a contradiction. 
In contrast to $\mu_1$, the measure $\mu_2$ does not make the problem easier.

When $T$ is $\omega(1)$, suppose that $e^{\ap}(B_n,T,p)$ is $o(T^{-1})$.
Let $n$ be sufficiently large. 
Then there exists a function $g$ with $g(T)=\omega(1)$ 
such that $e^{\ap}(B_n,T,p)\le (Tg(T))^{-1}$. Set
$\e=(Tg(T))^{-1}$ and observe that  $\e = o(1)$, as the second
statement of the theorem requires. This leads us to conclude
that $T$ must be
$\Omega(Tg(T))$ and, therefore, we get a contradiction. \hfill\slug

\vskip 1pc
For $\mu_2$, Theorem 3.3 and the results of \cite{brassard} 
and \cite{kwas} (for the worst
probabilistic error of QS) imply that QS is an asymptotically optimal
error and query algorithm.
Hence, in terms of error and number of necessary
queries, computing the Boolean mean on the average with
$\mu_2$ is as difficult as in the worst probabilistic case.

We end this section by extending our results to $\Delta$-approximate count. 
We present three corollaries. We omit their proofs since they are immediate 
from the corresponding theorems above.

\begin{cor} 
Consider $\Delta$-approximate count.
Let $I\subseteq \{0,\dots,n\}$ be a set of indices, such that
its cardinality $|I|$, as a function of $n$, is $\omega(1)$, and
$k(n-k)$ is $\Theta(n^2)$ for every $k\in I$. Assume that $\mu$ is a
probability measure on $B_n$ such that 
$$\mu(X)=\Omega(|I|^{-1})\frac 1{{n\choose |X|}}, \quad
{\rm for\ every\ } |X|\in I,\quad X\in B_n.$$
Then for any $\Delta>0$ of order $o(|I|)$, $e^{\ap}_1(B_n,T,p)\le\Delta$
implies that $T= \Omega(\min(n/ \Delta, n))$.
\end{cor}

\begin{cor}
Consider $\Delta$-approximate count 
and the average probabilistic error of a quantum 
algorithm with respect to $\mu_1$. The following two statements hold.
\begin{enumerate}
\item Let $T$ be $o(n)$. Then the error of any quantum algorithm 
with $T$ queries satisfies
$$e^{\ap}_1(B_n,T,p) = \Omega\left( \min\{ n^{1/2}, n/T \}\right).$$
\item Let $\Delta>0$ be $o(n^{1/2})$. Then
the number of queries $T(\Delta)$ for error at most $\Delta$
satisfies
$$T(\Delta)=\Omega\left(\min\{ n/ \Delta, n \}\right).$$
\end{enumerate}
\end{cor}

\begin{cor}
Consider $\Delta$-approximate count 
and the average probabilistic error of a quantum 
algorithm with respect to $\mu_2$. The following two statements hold.
\begin{enumerate}
\item Let $T$ be $o(n)$. Then the error of any quantum algorithm 
with $T$ queries satisfies
$$e^{\ap}_1(B_n,T,p) = \Omega\left( n/ T \right).$$
\item Let $\Delta>0$ be $o(n)$.
Then the number of queries $T(\Delta)$ for error at most $\Delta$
satisfies
$$T(\Delta)= \Omega\left(\min\{ n/ \Delta, n \}\right).$$
\end{enumerate}
\end{cor}

\vskip2pc
\section{Worst expected error}
\vskip 1pc

In this section we consider quantum algorithms with a worst expected error
criterion. We show query lower bounds for any quantum algorithm computing
the Boolean mean and for any quantum algorithm
computing a $\Delta$-approximate count.

\begin{thm} Consider any algorithm that computes the Boolean mean
with worst expected error satisfying $e^{\we}(q,B_n,T)\le \e$, for a fixed
$q\in [1,\infty)$. Then
the number of queries of this algorithm satisfies 
$$T(\e)= \Omega(\min\{ \e^{-1}, n\}).$$
\end{thm}

\noindent{\bf Proof:} Consider
$e^{\we}(q, B_n,T)\le \e$ and raise both sides to the power $q$ and
multiply them by $n^q$. Then set 
$\Delta = n\e$ and
$\Delta(X,j)=n |a_X-\hat a_X(j)|$ to obtain
$$\Delta^q \ge \sum_{j=0}^{M-1} \Delta(X,j)^q p_X(j), \quad \forall \; 
X\in B_n.$$

For any $\d>0$ using the Markov inequality we have
\begin{equation*}
\Delta^q 
\ge \d^q \sum_{\Delta(X,j) \ge \d} p_X(j), \quad\forall\; X\in B_n.
\end{equation*}

Choose a number $a>2$ and set $\d = a \D$ and $p= 1 - a^{-q}$, 
$p\in ( \tfrac 12 ,1)$.
Define 
$p_{\loss}(X) = \sum_{\Delta(X,j) \ge \d} p_X(j)$ for $X\in B_n$. Then
$$1-p = a^{-q} = \frac {\Delta^q}{\d^q} \ge p_{\loss}(X).$$
This implies that $p_{\win}(X)= 1-p_{\loss}(X)\ge p> \tfrac 12$, 
$\forall \; X\in B_n$.

Then there exist outcomes $j$ for which  $\d >\Delta(X,j)$
with probability 
$$p_{\win}(X)= \sum_{ n |a_X - \hat a_X(j)| < \d  }p_X(j)
\ge p >1/2, \quad \forall\; X\in B_n.$$
Hence, the probabilistic error of
$\d$-approximate count is
$e_1(X,T,p)< \d$, for all $X\in B_n$. Now take any $X$ such that $|X|
(n-|X|)=\Theta(n^2)$ and use the results of \cite{nyakWu} to see that
the number of necessary queries is 
$$\Omega\left(\sqrt{\frac n{\d}}+ \frac {\sqrt{|X|(n-|X|)}}{\d} \right).$$
Therefore, the number of queries satisfies
$$\Omega\left( \min\{ \e^{-1}, n\}\right).$$
\hfill\slug

It has been recently shown in \cite{hkw} that if one repeats $2(\lceil q\rceil 
+1)$ times the QS algorithm of Brassard et al. (with $T$ queries) 
then the median of the outputs has worst expected error 
of order $O(T^{-1})$. Using the theorem above we conclude that this is
an asymptotically optimal algorithm.

The following query lower bound for $\Delta$-approximate count is a
direct consequence of Theorem 4.1.

\begin{cor} Consider any algorithm that computes a $\Delta$-approximate count
with worst expected error satisfying $e^{\we}_1(q,B_n,T)\le \D$,
for fixed $q\in [1,\infty)$. Then
the number of queries of this algorithm satisfies 
$$T(\D)= \Omega(\min\{ n/\Delta, n\}).$$
\end{cor}

\vskip 2pc
\section{Average expected error}
\vskip 1pc

In this section we consider the average expected error of quantum algorithms. 
For brevity we call this the average expected setting.
Recall that we are considering the average with respect to 
a probability measure on the set of inputs $B_n$ and for each of the inputs
we consider the expected error of the quantum algorithm with respect to
all possible oucomes. 
We show query lower bounds for and quantum algorithm
computing the Boolean mean and for any quantum algorithm computing a 
$\D$-approximate count.

We deal only with the measures of Theorem 3.1 since $\mu_1$ and $\mu_2$ are
special cases that can be dealt with in the same way. In fact, Theorem 3.1
holds for the average expected error as well. The proof is based on that
of Theorem 3.1.

\begin{thm}
Consider the approximation of the Boolean mean.
Let $I\subseteq \{0,\dots,n\}$ be a set of consecutive indices, such that
its cardinality $|I|$, as a function of $n$, is $\omega(1)$. Assume that
$k(n-k)$ is $\Theta(n^2)$ for every $k\in I$. Let $\mu$ be a
probability measure on $B_n$ such that 
$$\mu(X)=\Omega(|I|^{-1})\frac 1{{n\choose |X|}}, \quad
{\rm for\ every\ }  X\in B_n {\rm \ with\ }|X|\in I.$$
Consider a fixed $q\in [1,\infty)$.
Then for any $\e>0$ of order $o(|I|n^{-1})$, the condition 
$e^{\avex}(q,B_n,T)\le\e$
implies that $T$ must be $\Omega(\min(\e^{-1}, n))$.
\end{thm}

\noindent{\bf Proof:} 
The proof is almost identical to that of Theorem 3.1 and we will only point
out the differences.

In particular,
for $1\leq q< \infty$ consider a quantum algorithm with average expected 
error at most $\e$, i.e.,
$$\e\ge e^{\avex}(B_n,T)=\sum_{X\in B_n}e^{\ex}(X,T)\mu(X).$$
We follow the first part of the proof of Theorem 3.1 replacing $e(X,T,p)$
by $e^{\ex}(X,T)$ and redefining the rest of
the quantities accordingly.
After the two applications of the Markov inequality we 
know that the number of strings $X$ for which
$\d(X,k)<2m\D$, $|X|=k$, is $\tilde n_-\ge (1-m^{-1}){n\choose k}$, 
and $m\ge 2$. Recall that $\d(X,k)=n e^{\ex}(X,T)$.

Using the Markov inequality as in Theorem 4.1
to derive the probabilistic error from the expected error,
we conclude that the probabilistic error of approximate
count satisfies $e_1(X,T,p) < 2ma\D$, with probability $p\ge 1-a^{-q}>1/2$ 
for a chosen $a>2$.

Now we return to the proof of Theorem 3.1. We have that there exist $k_1, k_2
\in J$ that are at least $4ma\D$ apart whose distance does not exceed
$O(\D)$, i.e., $|k_1 - k_2|\ge 4ma\D$ and $|k_1 - k_2|= O(\D)$, and 
$e_1(X,T,p)\le 2ma\D$, $|X|= k_1$, or $k_2$.

We use the original algorithm that approximates the mean 
to derive a new algorithm 
that approximates the partial Boolean function 
\begin{equation*}
f_{k_1,k_2}(X)=\left\{ 
\begin{array}{lcl}
1 &\quad &{\rm if\ } |X|=k_1 \\
0 &\quad &{\rm if\ } |X|=k_2.
\end{array}
\right.
\end{equation*}
where, without loss of generality, we can assume that $k_1>k_2$.

The description of the new algorithm $A$ is as follows: 
On input $X$, where $|X|=k_1$ or $k_2$, we run
the algorithm that approximates the mean and if the value of the result
$\hat a_X(j)$ satisfies $|k_1-n\hat a_X(j)|<2ma\Delta$ then the new algorithm
outputs $1$. It outputs $0$ otherwise.

There is one more difference between this proof and the proof of Theorem 3.1.
It concerns the derivation of the success/failure probability of 
the new algorithm and we explain this difference below.

Let's look at the success probability $\Prob\{ A(X)=f_{k_1,k_2}(X)\}$
of the new algorithm 
for the different inputs for which $f_{k_1,k_2}$ is defined. 
If $|X|=k_1$ we have
\begin{eqnarray*}
\sum_{|X|=k_1} \Prob\{ A(X)=f_{k_1,k_2}(X)\}
&=& \sum_{|X|=k_1} \Prob\{ A(X)=1 \} \\
&\ge&  \sum_{|X|=k_1,\; \d(X,k) < 2m\Delta} \Prob\{ A(X)=1 \} \\
&=& \sum_{|X|=k_1,\; \d(X,k)<2m\Delta} \Prob\{ |k_1-n\hat a_X(j)|<2ma\Delta \}\\
&\ge& (1-m^{-1}) {n \choose k_1} (1-a^{-q}),
\end{eqnarray*}
because we saw that when the expected error satisfies
$\d(X,k)<2m\Delta$ then this implies that 
the probabilistic error satisfies
$e_1(X,T,p)< 2ma\Delta$ with probability
$p\ge 1-a^{-q}>1/2$. The fact that $\tilde n_-\ge (1-m^{-1}) 
{n \choose k_1}$ yields the final inequality.
Therefore, the probability that algorithm $A$ fails on any input $X$
for which $|X|=k_1$ satisfies
\begin{equation*}
\sum_{|X|=k_1}\Prob \{ A(X)\ne f_{k_1,k_2}(X) \} = 
\sum_{|X|=k_1}\Prob \{ A(X)=0\} \le {n\choose k_1} \left(a^{-q}+
\frac {1-a^{-q}}m\right).
\end{equation*}
Let $c_2=a^{-q}+(1-a^{-q})/m$, where $a>2$. 
Just like in the proof of Theorem 3.1, we choose $m$ in a way
that $c_2< \tfrac 12$. This leads us to the equivalent of (\ref{eq:probloss1})
of Theorem 3.1.

In the same way we derive the equation concerning the probability of failure
of the new algorithm on input $|X|=k_2$ which corresponds to equation
(\ref{eq:probloss2}) of Theorem 3.1.

The remaining steps are identical to those of Theorem 3.1 and complete the
proof. \hfill\slug

\vskip 1pc

Theorem 5.1 shows that QS algorithm with repetitions \cite{hkw} is 
asymptotically optimal in the average expected case when the required accuracy 
is high. In fact, the query lower bounds of section 3 that depend either on 
Theorem 3.1 directly or have been derived through as similar proof technique
extend to the average expected and we have seen how this can be accomplished
in the proof of Theorem 5.1.

The following corollary 
for $\D$-approximate count in the average expected case is immediate.

\begin{cor} 
Consider $\Delta$-approximate count.
Let $I\subseteq \{0,\dots,n\}$ be a set of consecutive indices, such that
its cardinality $|I|$, as a function of $n$, is $\omega(1)$, and
$k(n-k)$ is $\Theta(n^2)$ for every $k\in I$. Assume that $\mu$ is a
probability measure on $B_n$ such that 
$$\mu(X)=\Omega(|I|^{-1})\frac 1{{n\choose |X|}}, \quad
{\rm for\ every\ } |X|\in I,\quad X\in B_n.$$
Consider a fixed $q\in [1,\infty)$.
Then for any $\Delta>0$ of order $o(|I|)$, $e^{\avex}_1(q,B_n,T)\le\Delta$
implies that $T= \Omega(\min(n/ \Delta, n))$.
\end{cor}

\vskip 2pc
\section*{Acknowledgements}
\vskip 1pc

I thank P. Jaksch, J. Traub, A. Werschulz and H. Wo\'zniakowski 
for their comments 
and suggestions that significantly improved this paper.

\vskip 2pc
\section{Appendix}
\vskip 1pc

\begin{lem} For $n\in\naturals$ and $1\le c\le \sqrt n/6$ we have
$${n \choose {n/2 \pm c\sqrt{n}}} > e^{-6c^2-2} \frac{2^n}{\sqrt{2\pi n}}.$$
\end{lem}

\noindent{\bf Proof:} From Stirling's formula \cite[p. 257]{abram} we have
$$
 n! = \sqrt{2\pi } n^{n+1/2} e^{-n+\theta / (12 n)},\quad 0<\theta < 1. 
$$
Thus,
$$
n! \le e\sqrt{2\pi n} ( n/e )^n,
$$
and
$$
n! \ge \sqrt{2\pi n} (n/e)^n.
$$
Therefore,
\begin{eqnarray*}
(n/2 + c\sqrt n)! &\le& e \left( \frac{n+2c\sqrt n}{2e}\right)^{n/2+c\sqrt n}
\sqrt{2\pi(n/2+c\sqrt n)} \\
&<& e  \left( \frac{n+2c\sqrt n}{2e}\right)^{n/2+c\sqrt n}
\sqrt{2\pi n},
\end{eqnarray*}
and
\begin{eqnarray*}
(n/2 - c\sqrt n)! &\le& e \left( \frac{n-2c\sqrt n}{2e}\right)^{n/2-c\sqrt n}
\sqrt{2\pi(n/2-c\sqrt n)} \\
&<& e  \left( \frac{n-2c\sqrt n}{2e}\right)^{n/2-c\sqrt n}\sqrt{2\pi n}.
\end{eqnarray*}

From the inequalities above we obtain
\begin{eqnarray*}
\frac{n!}{(n/2+c\sqrt{n})! (n/2-c\sqrt{n})!} &>& 
\frac{2^n n^n}{e^2\sqrt{2\pi n} (n+2c\sqrt n)^{n/2+c\sqrt n}
(n-2c\sqrt n)^{n/2-c\sqrt n}} \\
&=& 
\frac{2^n n^n}{e^2\sqrt{2\pi n} (n+2c\sqrt n)^{n/2} (n-2c\sqrt n)^{n/2}} 
\left( \frac{n-2c\sqrt n}{n+2c\sqrt n}\right)^{c\sqrt n} \\
&=& 
\frac{2^n n^n}{e^2\sqrt{2\pi n} (n^2-4c^2 n)^{n/2} }
\left( \frac{\sqrt n-2c}{\sqrt n+2c}\right)^{c\sqrt n} \\
&>&
\frac{2^n n^n}{e^2\sqrt{2\pi n} n^n} 
\left( \frac{\sqrt n-2c}{\sqrt n+2c}\right)^{c\sqrt n} \\
&=&
\frac{2^n}{e^2\sqrt{2\pi n}  }
\left( \frac{\sqrt n-2c}{\sqrt n+2c}\right)^{c\sqrt n} 
\end{eqnarray*}
Using 
$$\left( \frac{\sqrt{n}+2c}{\sqrt{n}-2c}\right)^{\sqrt n - 2c}=
\left( 1 + \frac{4c}{\sqrt n - 2c}\right)^{\sqrt n -2c} 
<e^{4c},$$
we obtain that
$$\left( \frac{\sqrt{n}+2c}{\sqrt{n}-2c}\right)^{c\sqrt n }
<e^{4c^2} \left( \frac{\sqrt{n}+2c}{\sqrt{n}-2c}\right)^{2c^2}\le 
e^{6c^2}.$$
Thus,
$$
\frac{n!}{(n/2+c\sqrt{n})! (n/2-c\sqrt{n})!} > 
\frac {2^n}{e^{6c^2+2}\sqrt{2\pi n}}.  $$
\hfill\slug

\vskip 2pc

\end{document}